\documentclass[12pt]{article}
\usepackage{bm,amsmath,amssymb,graphicx}

\begin{document}
\pagenumbering{arabic}
\begin{titlepage}

\title{Solar system tests for massive conformal gravity}

\author{F. F. Faria$\,^{*}$ \\
Centro de Ci\^encias da Natureza, \\
Universidade Estadual do Piau\'i, \\ 
64002-150 Teresina, PI, Brazil}

\date{}
\maketitle

\begin{abstract}
We find the linearized gravitational field of a static spherically 
symmetric mass distribution in massive conformal gravity and test 
it with some solar system experiments. The result is that the theory 
agrees with the general relativistic observations in the solar 
system for a determined lower bound on the graviton mass.
\end{abstract}

\thispagestyle{empty}
\vfill
 \par
\bigskip
\noindent * felfrafar@hotmail.com \par
\end{titlepage}
\newpage


\section{Introduction}
\label{sec1}


It is well known that general relativity (GR) is consistent with all the solar 
system experimental tests. However, in order 
to explain the galaxies rotation curves and the deflection of light by 
galaxies and clusters of galaxies, GR requires the existence of large 
amounts of dark matter whose nature is still unknown. A number of 
alternative theories of gravity (see, e.g., 
\cite{Weyl,Kaluza,Brans,Bergmann,Milgrom,Horava}) have been proposed 
to solve this and other GR problems but none have been completely successful 
so far. 

One of the alternative theories of gravity that solves 
the dark matter problem is conformal gravity (CG) \cite{Mann1}. However, in 
order to be compatible with the solar system observations, the theory needs 
to couple to a non-positive source with a highly singular structure \cite{Mann2}, 
which is not physical until proven otherwise. By considering that the conformal 
symmetry is likely to be important at the quantum level \cite{Hooft}, it is 
interesting to check if this problem in the CG description of the solar system 
phenomenology also happens in other conformally invariant theories of gravity. 
One of the most promising of such theories is the so called massive conformal 
gravity (MCG), which is a conformally invariant theory of gravity in which the 
gravitational action is the sum of the fourth order derivative Weyl action with 
the second order derivative Einstein-Hilbert action conformally coupled to a 
scalar field \cite{Faria1}.
 
So far it has been shown that MCG is free of the vDVZ discontinuity 
\cite{Faria2} and can reproduce the orbit of binaries 
by the emission of gravitational waves \cite{Faria3}. In addition, the 
theory is a power-counting renormalizable and unitary quantum theory of 
gravity \cite{Faria4,Faria5,Faria6}, and its cosmology describes the late 
universe without the cosmological constant problem \cite{Faria7}. Here we aim 
to check if MCG is consistent with some solar system experiments, which are the 
basic tests that any alternative theory of gravity must pass.
In Sec. \ref{sec2}, we describe the classical MCG equations. In Sec. 
\ref{sec3}, we derive the linearized MCG gravitational field of a static 
spherically symmetric massive body. In Secs. \ref{sec4}, \ref{sec5} and 
\ref{sec6}, we analyze if the theory is consistent with the deflection of 
light by the Sun, the radar echo delay and the perihelion precession of 
Mercury, respectively. Finally, conclusions are given in Sec. \ref{sec7}.     


\section{Massive conformal gravity}
\label{sec2}


Let us consider the total MCG action, which is given by\footnote{This action 
is obtained from the action of Ref. \cite{Faria2} by rescaling $\varphi 
\rightarrow \left(\sqrt{32\pi G/3}\right)\varphi$ and considering 
 $m = \sqrt{3/64\pi G\alpha}$.} \cite{Faria2}
\begin{equation}
S = \int{d^{4}x} \, \sqrt{-g}\bigg[ \varphi^{2}R 
+ 6 \partial^{\mu}\varphi\partial_{\mu}\varphi 
- \frac{1}{2\alpha^2} C^{\alpha\beta\mu\nu}C_{\alpha\beta\mu\nu} \bigg] 
+ \int{d^{4}x\mathcal{L}_{m}},
\label{1}
\end{equation}
where $\varphi$ is a scalar field called 
dilaton, $\alpha$ is a dimensionless constant, 
\begin{equation}
C^{\alpha\beta\mu\nu}C_{\alpha\beta\mu\nu} = R^{\alpha\beta\mu\nu}
R_{\alpha\beta\mu\nu} - 2R^{\mu\nu}R_{\mu\nu} + \frac{1}{3}R^{2}
\label{2}
\end{equation}
is the Weyl tensor squared,
$R^{\alpha}\,\!\!_{\mu\beta\nu} 
= \partial_{\beta}\Gamma^{\alpha}_{\mu\nu} + \cdots$ is the Riemann tensor, 
$R_{\mu\nu} = R^{\alpha}\,\!\!_{\mu\alpha\nu}$ is the Ricci tensor, 
$R = g^{\mu\nu}R_{\mu\nu}$ is the scalar curvature, and 
$\mathcal{L}_{m} = \mathcal{L}_{m}(g_{\mu\nu},\Psi)$ is the 
conformally invariant Lagrangian 
density of the matter field $\Psi$. It is worth noting 
that besides being invariant under coordinate transformations, the action 
(\ref{1}) is also invariant under the conformal transformations
\begin{equation}
\tilde{\Phi} = \Omega(x)^{-\Delta_{\Phi}}\Phi,
\label{3}
\end{equation}
where $\Omega(x)$  is an arbitrary function of the spacetime coordinates, 
and $\Delta_{\Phi}$ is the scaling dimension of the field $\Phi$, whose 
values are $-2$ for the metric field, $0$ for gauge bosons, $1$ for 
scalar fields, and $3/2$ for fermions.  

Varying the action (\ref{1}) with respect to $g^{\mu\nu}$ and $\varphi$, 
we obtain the MCG field equations \cite{Faria2}
\begin{equation}
\varphi^{2}G_{\mu\nu} +  6 \partial_{\mu}\varphi\partial_{\nu}\varphi 
- 3g_{\mu\nu}\partial^{\rho}\varphi\partial_{\rho}\varphi + g_{\mu\nu} 
\nabla^{\rho}\nabla_{\rho} \varphi^{2} 
- \nabla_{\mu}\nabla_{\nu} \varphi^{2}  - \alpha^{-2} W_{\mu\nu} 
= \frac{1}{2}T_{\mu\nu},
\label{4}
\end{equation}
\begin{equation}
\left(\nabla^{\mu}\nabla_{\mu} - \frac{1}{6}R \right) \varphi = 0,
\label{5}
\end{equation}
where
\begin{eqnarray}
W_{\mu\nu} &=& \nabla^{\rho}\nabla_{\rho}R_{\mu\nu} 
- \frac{1}{3}\nabla_{\mu}\nabla_{\nu}R  -\frac{1}{6}g_{\mu\nu}\nabla^{\rho}
\nabla_{\rho}R + 2R^{\rho\sigma}R_{\mu\rho\nu\sigma} 
-\frac{1}{2}g_{\mu\nu}R^{\rho\sigma}R_{\rho\sigma}  \nonumber \\ &&
- \frac{2}{3}RR_{\mu\nu}  + \frac{1}{6}g_{\mu\nu}R^2
\label{6}
\end{eqnarray}
is the Bach tensor,
\begin{equation}
G_{\mu\nu} = R_{\mu\nu} - \frac{1}{2}g_{\mu\nu}R
\label{7}
\end{equation}
is the Einstein tensor,
\begin{equation}
\nabla^{\rho}\nabla_{\rho} \varphi = 
\frac{1}{\sqrt{-g}}\partial^{\rho}\left( \sqrt{-g} \partial_{\rho}
\varphi \right)
\label{8}
\end{equation} 
is the generally covariant d'Alembertian for a scalar field, and
\begin{equation}
T_{\mu\nu} = \frac{2}{\sqrt{-g}} \frac{\delta \mathcal{L}_{m}}
{\delta g^{\mu\nu}}
\label{9}
\end{equation}
is the matter energy-momentum tensor. 

By considering that, at scales below the Planck scale, the dilaton field acquires 
a spontaneously broken constant vacuum expectation value $\varphi_{0}$ \cite{Matsuo},  
we find that the MCG field equations (\ref{4}) and (\ref{5}) become 
\begin{equation}
\varphi_{0}^{2}G_{\mu\nu} - \alpha^{-2} W_{\mu\nu} = \frac{1}{2}T_{\mu\nu},
\label{10}
\end{equation}
\begin{equation}
R = 0.
\label{11}
\end{equation} 
In addition, for $\varphi = \varphi_0$, the MCG line element 
$ds^2 = \left(\varphi/\varphi_{0}\right)^{2}g_{\mu\nu}dx^{\mu}dx^{\nu}$  
reduces to the general relativistic line element  
\begin{equation}
ds^2 = g_{\mu\nu}dx^{\mu}dx^{\nu},
\label{12}
\end{equation}
and the MCG geodesic equation
\begin{equation}
\frac{d^{2}x^{\lambda}}{d\tau^2} + \Gamma^{\lambda}\,\!\!_{\mu\nu}
\frac{dx^{\mu}}{d\tau}\frac{dx^{\nu}}{d\tau} +\frac{1}{\varphi}
\frac{\partial\varphi}{\partial x^{\rho}} \left( g^{\lambda\rho} + 
\frac{dx^{\lambda}}{d\tau}\frac{dx^{\rho}}{d\tau}\right) = 0
\label{13}
\end{equation}
reduces to the general relativistic geodesic equation
\begin{equation}
\frac{d^{2}x^{\lambda}}{d\tau^2} + \Gamma^{\lambda}\,\!\!_{\mu\nu}
\frac{dx^{\mu}}{d\tau}\frac{dx^{\nu}}{d\tau} = 0,
\label{14}
\end{equation}
where
\begin{equation}
\Gamma^{\lambda}\,\!\!_{\mu\nu} = \frac{1}{2}g^{\lambda\rho}\left( 
\partial_{\mu}g_{\nu\rho} + \partial_{\nu}g_{\mu\rho} 
- \partial_{\rho}g_{\mu\nu} \right)
\label{15}
\end{equation}
is the Levi-Civita connection. The full classical content of MCG can be 
obtained from (\ref{10}), (\ref{11}), (\ref{12}) and (\ref{14}) without loss 
of generality.


\section{Linearized static solution}
\label{sec3}


In order to submit the theory to solar system tests we need to find 
static spherically symmetric solutions to the linear MCG field equations. 
The substitution of the linearized metric in isotropic coordinates
\begin{equation}
ds^{2} = - \left[ 1 + \frac{2V(r)}{c^2}\right]c^2dt^{2} + \left[1 
- \frac{2W(r)}{c^2}\right]\delta_{ij}dx^idx^j,
\label{16}
\end{equation}
and the MCG energy-momentum tensor of a point particle 
source with mass $M$ at rest at the origin (see Ref. \cite{Faria2})
\begin{equation}
T_{\mu\nu} = \left( \delta^{0}_{\mu}\delta^{0}_{\nu} 
+ \frac{1}{4}\eta_{\mu\nu}\right)Mc^2 \delta^3(\textbf{r}),
\label{17}
\end{equation}
into (\ref{10}) and (\ref{11}) gives
\begin{equation} 
\varphi_{0}^{2}\left(\nabla^{2}W\right) 
-\frac{1}{3\alpha^2}\left(\nabla^4 V+\nabla^4 W\right) 
= \frac{3Mc^{4}}{16}\delta^3(\textbf{r}),
\label{18}
\end{equation}
\begin{equation}
 \nabla^{2}V - 2\nabla^{2}W = 0,
\label{19}
\end{equation}
where $\nabla^{2}$ is the Laplacian.

By using (\ref{19}), we can write (\ref{18}) 
in the form
\begin{equation}
\left[ \left(\frac{\hbar}{mc}\right)^2\nabla^2 - 1 \right]\nabla^2 V = 
-\frac{3Mc^4}{8\varphi_{0}^{2}}\delta^3(\textbf{r}),
\label{20}
\end{equation}
where $m=\alpha\varphi_{0}\hbar/c$ is the mass of a massive 
spin-$2$ field (ghost) with negative energy that appears in the 
theory in addition to the usual massless spin-$2$ field (graviton) 
with positive energy \cite{Faria4}. The general solution to (\ref{20}) is 
given by
\begin{eqnarray}
V(r) &=& -\frac{3Mc^4}{8\varphi_{0}^{2}}\int \frac{d^{3}\textbf{p}}{(2\pi)^4}
\frac{(mc/\hbar)^2 e^{i\textbf{p}\cdot \textbf{x}}}{\textbf{p}^{2}
\left[\textbf{p}^{2} + (mc/\hbar)^2\right]} \nonumber \\ &=& 
- \frac{3Mc^4}{8\varphi_{0}^{2}}\left(\frac{1- e^{-\frac{mc}{\hbar} r}}
{4\pi r}\right).
\label{21}
\end{eqnarray}
In order to this solution agree with the Newtonian potential 
in the limit where $m$ tend to infinity, we must choose 
$\varphi_{0}^{2}= 3c^4/32\pi G$, which gives
\begin{equation}
V(r) = -\frac{GM}{r}\left( 1- e^{-\frac{mc}{\hbar} r}\right).
\label{22}
\end{equation}
It is not difficult to see that (\ref{22}) tends to the finite value 
$GMmc/\hbar$ at the origin, which is a necessary condition for 
the renormalizability of the theory \cite{Gia}.

Solving (\ref{20}) in vacuum, we obtain
\begin{equation}
V(r) = C_1 + \frac{C_2}{r} + \frac{C_3 e^{-\frac{mc}{\hbar}r}}{r} 
+ \frac{C_4 e^{\frac{mc}{\hbar}r}}{r}, 
\label{23}
\end{equation}
where $C_1$, $C_2$, $C_3$ and $C_4$ are arbitrary constants. 
The substitution of (\ref{23}) into (\ref{19}) then gives
\begin{equation}
W(r) = \frac{C_2}{r} + \frac{1}{2}\frac{C_3 e^{-\frac{mc}{\hbar}r}}{r}
 + \frac{1}{2}\frac{C_4 e^{\frac{mc}{\hbar}r}}{r}.
\label{24}
\end{equation}
The comparison of (\ref{23}) with (\ref{22}) requires that $C_1 = C_4 = 0$, 
$C_2 = -GM$ and $C_3 = GM$. Finally, using these values in (\ref{24}), we find
\begin{equation}
W(r) = -\frac{GM}{r}\left( 1- \frac{1}{2}e^{-\frac{mc}{\hbar}r}\right),
\label{25}
\end{equation}
which diverges to infinity as we approaches the origin. Since this 
singular behavior extends as well to the linearized curvature invariants 
$R^{\mu\nu}R_{\mu\nu}$ and $R^{\alpha\beta\mu\nu}R_{\alpha\beta\mu\nu}$, 
the origin is an actual singularity. However, it is well known that the 
linearized approximation breaks down in the vicinity of the origin and thus 
we must use the full nonlinear equations of the theory in order to analyze 
the singularity at the origin, which is far beyond the scope of this paper.

The presence of the ghost can lead to instabilities in the classical solutions 
of the theory. In order to analyze the stability of (\ref{16}), we must 
consider the linearization of (\ref{10}) and (\ref{11}) in vacuum about the 
perturbed potentials 
\begin{equation}
\tilde{V}(r,t) = V(r) + \delta V(r,t),
\label{26}
\end{equation}
\begin{equation}
\tilde{W}(r,t) = W(r) + \delta W(r,t),
\label{27}
\end{equation}
where $V(r)$ and $W(r)$ are given by (\ref{22}) and (\ref{25}), respectively, 
and $\delta V$ and $\delta W$ are small perturbations. In this case, using
(\ref{18}) and (\ref{19}) in vacuum, we find 
\begin{equation} 
\left(\frac{mc}{\hbar}\right)^2\Box \delta W 
-\frac{1}{3}\left(\Box^2 \delta V + \Box^2 \delta W\right) 
= 0,
\label{28}
\end{equation}
\begin{equation}
 \Box \delta V - 2\Box \delta W = 0,
\label{29}
\end{equation}
where $\Box$ is the d'Alembertian. 

We can solve (\ref{28}) and (\ref{29}) in the same way that we 
solved (\ref{18}) and (\ref{19}). The result is given by the spherical waves
\begin{equation}
\delta V  = \frac{C_1}{kr}\cos(kr-\omega_{k} t) 
+ \frac{C_2}{qr}\cos(qr-\omega_{q} t),
\label{30}
\end{equation}
\begin{equation}
\delta W  = \frac{C_1}{kr}\cos(kr-\omega_{k} t) 
+ \frac{C_2}{2qr}\cos(qr-\omega_{q} t),
\label{31}
\end{equation}
where $C_1$ and $C_2$ are constants, $\omega_{k} = kc$ 
and $\omega_{q} = \sqrt{(qc)^2 + m^2c^4/\hbar^2}$. Since (\ref{30}) and 
(\ref{31}) do not grow unboundedly with time, the static spherically symmetric 
solutions to the linear MCG field equations are stable.  

Before proceeding, it is worth noting that the 
MCG potentials (\ref{22}) and (\ref{25}) are not related with the CG 
potentials, which are given by\footnote{These potentials are 
obtained by coupling the theory with the usual positive delta source. The theory 
gives the general relativistic potentials only if coupled with a negative 
source involving much deeper singularities than 
the usual positive delta source, as stated in the introduction.} 
\cite{Flanagan}
\begin{equation}
V(r) = \frac{GM}{r}\left( 1 - \frac{4}{3}e^{-\frac{mc}{\hbar} r}
\right), \ \ \ \ \ W(r) = \frac{GM}{r}\left(1 - \frac{2}{3}
e^{-\frac{mc}{\hbar} r}\right).
\label{32}
\end{equation}
This is because MCG has one scalar field (dilaton)
coupled to the gravitational part of the theory and a
second scalar field (Higgs field) coupled to the matter part of the theory 
\cite{Faria7}, while CG has the Higgs field but not the dilaton \cite{Mann3}.


\section{Deflection of light by the Sun}
\label{sec4}


One observable effect that can be obtained from the linear approximation of 
MCG is the deflection of light by the gravitational field of the Sun. To 
describe such effect we must first find how light propagates in a 
gravitational field in the theory
\footnote{We do not use the parametrized post-Newtonian 
(PPN) formalism in the following sections because its application to massive 
scalar-tensor theories of gravity is ambiguous \cite{Dya}.}.
 
Taking the linear approximation $g_{\mu\nu} = \eta_{\mu\nu} + h_{\mu\nu}$ 
of the geodesic equation (\ref{14}) and multiplying by $m' d\tau$, we obtain
\begin{equation}
m' du^{\lambda} + \left( \partial_{\mu}h^{\lambda}\,\!\!_{\nu} 
- \frac{1}{2}\partial^{\lambda}h_{\mu\nu}\right)m' u^{\mu}dx^{\nu} = 0,
\label{33}
\end{equation}
where $m'$ is the mass of a test particle and $u^{\mu} = dx^{\mu}/d\tau$ is the 
test particle four-velocity, with $\tau$ being the proper time. By neglecting 
the terms of second order in $h_{\mu\nu}$, we can write (\ref{33}) as
\begin{equation}
dP^{\lambda} - \frac{1}{2}\partial^{\lambda}h_{\mu\nu}P^{\mu}dx^{\nu} = 0,
\label{34}
\end{equation}
where $P^{\mu} = m'(u^{\mu} + h^{\mu}\,\!\!_{\nu}u^{\nu})$ is the test particle 
momentum-energy four-vector. The form (\ref{34}) of the geodesic equation can 
be used to describe the trajectories of both particles and light waves in 
MCG.

The momentum-energy four-vector of a photon with frequency $\omega$ 
and wave vector $\vec{k} = (k_{x}, k_{y}, k_{z})$ is given by
\begin{equation}
P^{\mu} = \hbar k^{\mu},
\label{35}
\end{equation}
where
\begin{equation}
k^{\mu} = \left( \frac{\omega}{c}, \vec{k} \right) =  
\left( \frac{\omega}{c}, k_{x}, k_{y}, k_{z} \right)
\label{36}
\end{equation}
is the photon wave four-vector. Substituting (\ref{35}) into (\ref{34}), 
we obtain
\begin{equation}
dk^{\lambda} - \frac{1}{2}\partial^{\lambda}h_{\mu\nu}k^{\mu}
dx^{\nu} = 0.
\label{37}
\end{equation}

\begin{figure}[h]
 \centering
	\includegraphics[scale=0.5]{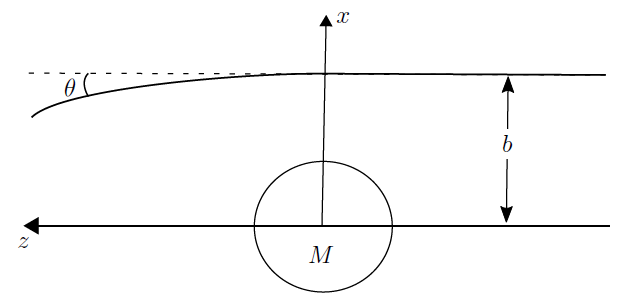}
	\caption{Path of a light ray  in the gravitational field of a static 
	spherically symmetric body of mass $M$.}
	\label{f1}
\end{figure}

For a photon passing by a static spherically symmetric massive body along 
the $z$-axis with impact parameter $b$ (see Fig. \ref{f1}), we have
\begin{equation}
k^{\mu} = \left(\frac{\omega}{c}, 0, 0, \frac{\omega}{c} \right),
\label{38}
\end{equation}
\begin{equation}
dx^{\mu} = (dz, 0, 0, dz).
\label{39}
\end{equation}
The insertion of (\ref{38}) and (\ref{39}) into (\ref{37}) then gives
\begin{equation}
dk^{\lambda} = \frac{\omega}{2c}\partial^{\lambda}\left(h_{00} + h_{03} 
+ h_{30} + h_{33} \right)dz.
\label{40}
\end{equation}
By substituting (\ref{16}) into (\ref{40}), we arrive at
\begin{equation}
dk_{x} = - \frac{\omega}{c^3}\frac{\partial }{\partial x}\left(V 
+ W\right)dz,
\label{41}
\end{equation}
so that the total change in $k_{x}$ is
\begin{equation}
\Delta k_{x} = - \frac{\omega}{c^3}\int^{\infty}_{-\infty}
\frac{\partial }{\partial x}\left(V + W\right)  \, dz,
\label{42}
\end{equation}
where the derivative is to be evaluated at the impact parameter of the light 
ray.

\begin{figure}[h]
 \centering
	\includegraphics[scale=0.6]{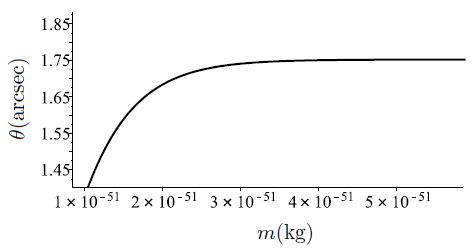}
	\caption{Deflection angle  as a function of mass for light
rays grazing the Sun in MCG.}
	\label{f2}
\end{figure}

The deflection angle for a light ray passing by a static spherically 
symmetric mass distribution is therefore \cite{Oha}
\begin{equation}
\theta = \left|\frac{\Delta k_{x}}{k_{z}}\right| 
= \frac{1}{c^2}\int^{\infty}_{-\infty}
\frac{\partial }{\partial x}\left(V + W\right)\Big|_{x=b} \, dz.
\label{43}
\end{equation}
Inserting (\ref{22}) and (\ref{25}) into (\ref{43}), we obtain the MCG 
deflection angle
\begin{equation}
\theta_{\textrm{MCG}} = \theta_{\textrm{GR}} 
- \frac{3GMb}{2c^2}\int^{\infty}_{-\infty}\left[ 
\frac{1+\frac{mc}{\hbar}\sqrt{b^2+z^2}}{(b^2+z^2)^{3/2}}\,e^{-\frac{mc}{\hbar}
\sqrt{b^2+z^2}}
 \right] dz,
\label{44}
\end{equation}
where
\begin{equation}
\theta_{\textrm{GR}} = \frac{2GMb}{c^2}\int^{\infty}_{-\infty}\left[ 
\frac{1}{(b^2+z^2)^{3/2}}\right] dz
\label{45}
\end{equation}
is the general relativistic deflection angle. Integrating (\ref{44}) 
numerically, with $M = M_{\odot} = 1.9891 \times 10^{30} \, \mbox{kg}$ 
and $b = 6.955 \times 10^{8} \, \mbox{m}$, we can find the evaluation 
of the solar MCG deflection angle for different values of $m$. The result 
depicted in Fig. \ref{f2} shows that we must have $m \gtrsim 3.5\times 
10^{-51} \, \mbox{kg}$ in order to the solar MCG deflection angle 
agrees with the measured solar GR deflection angle 
$\theta_{\textrm{GR}} \sim 1.75''$  \cite{Bruns}.

\section{Radar echo delay}
\label{sec5}

\begin{figure}[h]
 \centering
	\includegraphics[scale=0.5]{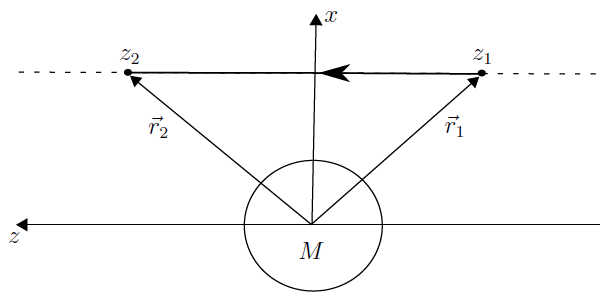}
	\caption{Path of a light signal between two points in the gravitational 
	field of a static spherically symmetric mass distribution.}
	\label{f3}
\end{figure}

Let us consider a light signal that moves along a straight line (parallel to 
the $z$-axis) connecting two points $z_{1}$ and $z_{2}$ in the gravitational 
field of a static spherically symmetric massive body (see Fig. \ref{f3}). It 
follows from (\ref{16}) and (\ref{40}) that the change in $k_{z}$ along the 
straight line is given by
\begin{equation}
dk_{z} = - \frac{\omega}{c^3}\frac{\partial }{\partial z}\left(V 
+ W\right)dz.
\label{46}
\end{equation}
Substituting (\ref{22}) and (\ref{25}) into (\ref{46}), and integrating, we 
find
\begin{eqnarray}
k_{z}(z) - \frac{\omega}{c} &=&
- \, \frac{2GM\omega}{c^3}\int^{z}_{\infty}\Bigg[\frac{z'}{(b^2+z'^2)^{3/2}}
- \frac{3}{4}\frac{z'e^{-\frac{mc}{\hbar}\sqrt{b^2+z'^2}} }{(b^2+z'^2)^{3/2}} 
\nonumber \\ && - \, \frac{3}{4}\frac{\frac{mc}{\hbar}z'
e^{-\frac{mc}{\hbar}\sqrt{b^2+z'^2}}}{\sqrt{b^2+z'^2}} \Bigg]  dz', 
\label{47}
\end{eqnarray}
where we used $k_{z}(\infty) = \omega/c$. 

\begin{figure}[h]
 \centering
	\includegraphics[scale=0.6]{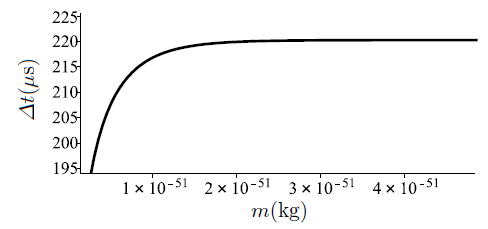}
	\caption{Time delay as a function of mass for light traveling between the 
Earth and Mercury in MCG.}
	\label{f4}
\end{figure}

The insertion of (\ref{47}) into the delay in the time it takes the light 
signal to travel from $z_{1}$ to $z_{2}$ and back \cite{Oha}
\begin{equation}
\Delta t = \frac{2}{\omega}\int_{z_{1}}^{z_{2}} \left[k_{z}(z) 
- \frac{\omega}{c}\right] \, dz
\label{48}
\end{equation}
gives
\begin{equation}
\Delta t_{\textrm{MCG}} = \Delta t_{\textrm{GR}} 
+ \frac{3GM}{c^3}\int_{z_{1}}^{z_{2}}\left[ \int^{z}_{\infty}
\left( \frac{1+\frac{mc}{\hbar}\sqrt{b^2+z'^2}}{(b^2+z'^2)^{3/2}}
\,z'e^{-\frac{mc}{\hbar}\sqrt{b^2+z'^2}}\right)  dz'\right] dz,
\label{49}
\end{equation}
where
\begin{equation}
\Delta t_{\textrm{GR}} = -\frac{4GM}{c^3}\int_{z_{1}}^{z_{2}}\left[ 
\int^{z}_{\infty}\left( \frac{z'}{(b^2+z'^2)^{3/2}}\right)  dz'\right] dz
\label{50}
\end{equation}
is the general relativistic time delay. For a light signal traveling between 
the Earth (at $z_{1} < 0$) and Mercury (at $z_{2} > 0$) on opposite sides of 
the Sun, we must set $M = M_{\odot}$, $b = 6.955 \times 10^{8} \, \mbox{m}$,  
$z_{1} = - \, 14.9 \times 10^{10} \, \mbox{m}$ and $z_{2} = 5.8 \times 10^{10} 
\, \mbox{m}$. Using these values, a numerical integration of (\ref{49}) gives 
the result shown in Fig. \ref{f4}, which leads to the conclusion that the MCG 
time delay is consistent with the observed general relativistic value 
$\Delta t_{\textrm{GR}} \sim 220 \, \mu\mbox{s}$ 
\cite{Bert} for $m \gtrsim 2\times 10^{-51} \, \mbox{kg}$.

\section{Perihelion precession of Mercury}
\label{sec6}

\begin{figure}[h]
 \centering
	\includegraphics[scale=0.55]{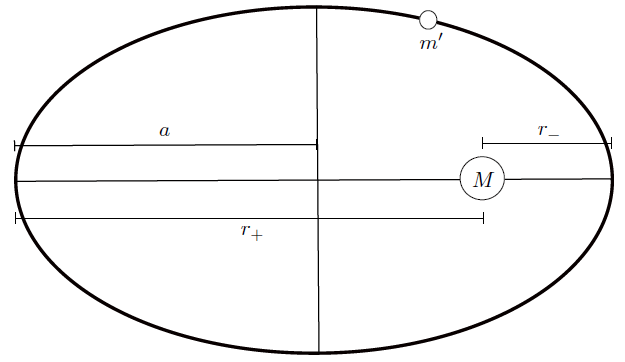}
	\caption{Test particle orbiting a static spherically symmetric mass 
	distribution.}
	\label{f5}
\end{figure}

Now we consider the motion of a test particle of mass $m'$ in orbit around a 
static spherically symmetric body of mass $M$ (see Fig. \ref{f5}). Using polar 
coordinates $r$ and $\phi$ in the orbital plane ($\theta = \pi/2$), 
we can write the total Newtonian energy of the system as
\begin{equation}
E = \frac{1}{2}\mu(\dot{r}^2+r^2\dot{\phi}^2) + U(r),
\label{51}
\end{equation}
where $\mu = m'M/(m'+ M)$ is the reduced mass and a dot denotes $d/dt$. 

Considering the conservation of the angular momentum 
\begin{equation}
J = \mu r^2\dot{\phi} = \mbox{constant},
\label{52}
\end{equation}
we rewrite (\ref{51}) in the form
\begin{equation}
E = \frac{1}{2}\mu\dot{r}^2 + \frac{J^2}{2\mu r^2} + U(r),
\label{53}
\end{equation}
which gives
\begin{equation}
\dot{r} = \frac{dr}{dt} = \sqrt{\frac{2}{\mu}\left[E - U(r)\right] 
- \frac{J^2}{\mu^2 r^2}}.
\label{54}
\end{equation}
By writing (\ref{52}) as $d\phi = Jdt/\mu r^2$, substituting $dt$ from 
(\ref{54}) and integrating, we obtain
\begin{equation}
\phi = \int{\frac{Jdr/r^2}{\sqrt{2\mu\left[E - U(r)\right]
- J^2/r^2}}} + \mbox{constant}.
\label{55}
\end{equation}

According to (\ref{55}), during the time in which $r$ varies from $r_{+}$ 
to $r_{-}$ and back, the radius vector turns through an angle
\begin{equation}
\Delta \phi = 2\int_{r_{-}}^{r_{+}}{\frac{Jdr/r^2}
{\sqrt{2\mu\left[E - U(r)\right]- J^2/r^2}}}.
\label{56}
\end{equation}
Substituting a potential energy of the type
\begin{equation}
U(r) = -\frac{GM\mu}{r} + \delta U(r),
\label{57}
\end{equation}
where $\delta U$ is a small correction to the Newtonian potential energy, 
into (\ref{56}) and expanding the integrand in powers of $\delta U$, we find that 
the zero-order term in the expansion gives $2\pi$, and the first-order 
term gives the precession of the orbit per revolution  \cite{Landau}
\begin{equation}
\delta \phi =  -\frac{2p}{GM\mu e} 
\int_{0}^{\pi}{r^2\frac{d}{dr}\left[\delta U(r)\right]\cos{\phi} 
\, d\phi},
\label{58}
\end{equation}
where
\begin{equation}
r = \frac{p}{(1+e\cos{\phi})},
\label{59}
\end{equation}
with $p = a(1-e^2)$ being the semilatus rectum, 
$a$ the semi-major axis and $e$ the eccentricity of the orbital ellipse. In 
terms of the variable $z = \cos{\phi}$, we can write (\ref{58}) in the more 
useful form \cite{Adkins}
\begin{equation}
\delta \phi =  -\frac{2p}{GM \mu e^2} 
\int_{-1}^{1}{\frac{z}{\sqrt{1-z^2}}\frac{d}{dz}\left[\delta U(z)\right]
\, dz}.
\label{60}
\end{equation}

In order to find the MCG potential energy, we substitute (\ref{16}) into the
normalization of the four-velocity
\begin{equation}
g_{\mu\nu}u^{\mu}u^{\nu} = -c^2,
\label{61}
\end{equation}
which, in polar coordinates in the orbital plane, gives
\begin{equation}
-\left( 1 + \frac{2V}{c^2}\right)
c^2 t'^2+\left( 1 - \frac{2W}{c^2}\right)
\left(r'^2+r^4\phi'^2 \right)= -c^2,
\label{62}
\end{equation}
where a prime denotes $d/d\tau$. Considering that the 
Lagrangian of the system is given by
\begin{equation}
\mathcal{L} = \frac{1}{2}\mu g_{\mu\nu} u^{\mu}u^{\nu} 
= \frac{1}{2}\mu\left[ -\left( 1 + \frac{2V}{c^2}\right)
c^2 t'^2+\left( 1 - \frac{2W}{c^2}\right)
\left(r'^2+r^4 \phi'^2 \right)\right],
\label{63}
\end{equation}
we find the total energy
\begin{equation} 
E_{\mathrm{tot}} = -\frac{\partial \mathcal{L}}{\partial t'} 
= \mu \left( 1 + \frac{2V}{c^2}\right)c^2 t'
\label{64}
\end{equation}
and the angular momentum 
\begin{equation} 
J = \frac{\partial \mathcal{L}}{\partial \phi'} 
= \mu \left( 1 - \frac{2W}{c^2}\right)r^4 \phi'.
\label{65}
\end{equation}

The substitution of $t'$ and $\phi'$ from (\ref{64}) 
and (\ref{65}) into (\ref{62}) gives
\begin{equation}
-\left( 1 + \frac{2V}{c^2}\right)^{-1}
\frac{E_{\mathrm{tot}}^2}{\mu^2c^2}+\left( 1 - \frac{2W}{c^2}\right)r'^2
+\left( 1 - \frac{2W}{c^2}\right)^{-1}\frac{J^2}{\mu^2r^2} = -c^2,
\label{66}
\end{equation}
which can be rewriting in the form
\begin{equation}
\frac{E_{\mathrm{tot}}^2}{c^2} = \left( 1 + \frac{2V}{c^2}\right)\mu^2
\left[\left( 1 - \frac{2W}{c^2}\right)r'^2
+\left( 1 - \frac{2W}{c^2}\right)^{-1}\frac{J^2}{\mu^2r^2} + c^2\right].
\label{67}
\end{equation}
By inserting (\ref{22}) and (\ref{25}) into (\ref{67}), keeping only the 
first-order kinetic term and the potential
terms up to second order in $1/r$, taking the Newtonian limit 
$d/d\tau \rightarrow d/dt$, and making some algebra, we obtain 
\begin{equation}
E = \frac{1}{2}\mu \dot{r}^2 + \frac{J^2}{2\mu r^2} -\frac{GM\mu}{r} 
+ \frac{GM\mu}{r}e^{-\frac{mc}{\hbar} r} 
- \frac{GMJ^2}{\mu c^2r^3}(1 - e^{-\frac{mc}{\hbar} r}),
\label{68}
\end{equation}
where
\begin{equation}
E = \frac{E_{\mathrm{tot}}^2 - \mu^2c^4}{2\mu c^2}
\label{69}
\end{equation}
is the usual Newtonian energy.

Comparing (\ref{68}) with (\ref{53}) and (\ref{57}), we can see that
\begin{equation}
\delta U(r) = \frac{GM\mu}{r}e^{-\frac{mc}{\hbar} r} 
- \frac{GMJ^2}{\mu c^2r^3}(1 - e^{-\frac{mc}{\hbar} r})
\label{70}
\end{equation}
for MCG. Finally, using $r = p/(1+ez)$ and $J^2=GM\mu^2 p$ in (\ref{70}), 
and substituting into (\ref{60}), we arrive at the MCG precession per 
revolution 
\begin{eqnarray}
\delta \phi_{\textrm{MCG}} &=& \delta \phi_{\textrm{GR}} -\frac{2p}{e} 
\int_{-1}^{1}\frac{z}{\sqrt{1-z^2}}\Bigg[\frac{1}{p}+ \frac{mc}{\hbar(1+ez)}
+ \frac{3GM(1+ez)^2}{c^2 p^2} \nonumber \\ && + \, \frac{GMm(1+ez)}{\hbar cp}
 \Bigg]\mbox{exp}\left(-\frac{mc}{\hbar}\frac{p}{1+ez}\right)  dz,
\label{71}
\end{eqnarray}
where
\begin{equation}
\delta \phi_{\textrm{GR}} = \frac{6GM}{e c^2 p} 
\int_{-1}^{1}{\frac{z}{\sqrt{1-z^2}}\left(1+ez\right)^2 \, dz} 
= \frac{6\pi GM}{c^2 p}
\label{72}
\end{equation}
is the general relativistic precession per revolution.

\begin{figure}[h]
 \centering
	\includegraphics[scale=0.6]{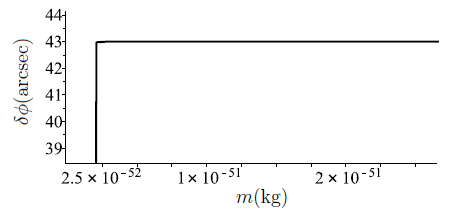}
	\caption{Precession per century for the orbit of 
	Mercury around the Sun as a function of mass in MCG.}
	\label{f6}
\end{figure}

Using $M = M_{\odot}$, $e = 0.2056$ and $p = 6.686 \times 10^{7} \, \mbox{km}$ 
in (\ref{71}), and integrating numerically, we find the MCG precession per 
century for the orbit of Mercury around the Sun, which is shown in 
Fig. \ref{f6}. Assuming that the measured precession for Mercury is 
$\sim 43''$ per century \cite{Park}, it follows that we must have 
$m \gtrsim 2.5 \times 10^{-52} \, \mbox{kg}$. 

\section{Final remarks}
\label{sec7}


In the present paper, we have compared the predictions of MCG with some solar 
system observations, namely, the deflection of light by the Sun,
the radar echo delay and the perihelion precession of Mercury. In particular, 
it was shown that the linear MCG predictions are consistent with these 
three solar system phenomena provided we have 
$m \gtrsim 3.5 \times 10^{-51} \, \mbox{kg}$. Despite 
this lower bound on the graviton mass is in agreement with the bound 
$m \gtrsim 10^{-38} \, \mbox{kg}$ imposed by Cavendish like experiments 
\cite{Adel} and the measured decrease 
of the orbital period of binary systems \cite{Faria3}, it makes the theory 
unable to explain galaxy rotation curves and the deflection of light by 
galaxies without dark matter. However, the conformal symmetry of the theory 
allows us to introduce an extra scalar field with zero vacuum
expectation value  in the matter part of the theory \cite{Faria8}. Although 
more studies on this are needed, this extra scalar may be a good candidate for 
dark matter. 

With the results obtained here, we have taken another important 
step towards confirming MCG as a serious candidate to solve the GR problems. 
Clearly, there are still many steps to be taken such as whether the 
theory is consistent with the early universe data or whether it solves the dark 
matter and singularity problems, among others. We will continue to work in the 
hope of overcoming some of these steps in the near future.


\end{document}